\documentclass[aps,prb,twocolumn,twoside,letterpaper,showpacs]{revtex4}

\usepackage{graphicx}
\usepackage{amssymb}
\usepackage{amsmath}

\def\cN{{\cal N}}
\def\vv{{\bf v}}
\def\vp{{\bf p}}

\def\vR{{\bf R}}

\newcommand{\D}{\mathfrak{D}}
\newcommand{\R}{\mathfrak{R}}

\begin{document}

\title{Proximity Effect in Normal Metal - High $T_c$ Superconductor Contacts}

\author{Tomas L$\mathrm{\ddot{o}}$fwander}

\affiliation{Department of Physics \& Astronomy, Northwestern University,
Evanston, IL 60208}

\date{\today}

\begin{abstract}

We study the proximity effect in good contacts between normal metals
and high $T_c$ ($d_{x^2-y^2}$-wave) superconductors. We present
theoretical results for the spatially dependent order parameter and
local density of states, including effects of impurity scattering in
the two sides, $s$-wave pairing interaction in the normal metal side
(attractive or repulsive), as well as subdominant $s$-wave paring in
the superconductor side. For the [100] orientation, a real combination
$d+s$ of the order parameters is always found. The spectral signatures
of the proximity effect in the normal metal includes a suppression of
the low-energy density of states and a finite energy peak
structure. These features are mainly due to the impurity
self-energies, which dominate over the effects of induced pair
potentials. For the [110] orientation, for moderate transparencies,
induction of a $d+is$ order parameter on the superconductor side,
leads to a proximity induced $is$ order parameter also in the normal
metal. The spectral signatures of this type of proximity effect are
potentially useful for probing time-reversal symmetry breaking at a
[110] interface.

\end{abstract}


\maketitle

\section{Introduction}

The proximity effect refers to a broad range of phenomena related to
the leakage of superconducting correlations into a normal metal in
contact with a superconductor. One way of viewing the proximity
effect, is to consider the decay of the electron pair wave function,
or off-diagonal retarded Green's function $f^R$, from the
superconductor side to the normal metal side, on the coherence length
scale.\cite{deu69}

The proximity effect in a normal metal in good contact with a
low-$T_c$ superconductor has been studied for a long
time\cite{deu69,gue96,mou01,gup04,Kieselmann,gol88,bel96,pil00} (see
also the review in Ref.~\onlinecite{bel99} and references therein). On
the other hand, the proximity effect in a normal metal in good contact
with a high-$T_c$ superconductor has not been considered
experimentally until very recently.\cite{koh03,sha04}

Sharoni {\it et al.}\cite{sha04} performed scanning tunneling
spectroscopy on gold coated $\mathrm{YBa_2Cu_3O_{7-\delta}}$ (YBCO),
and found a (pseudo-) gap on the gold side that decayed exponentially
with distance from the superconductor on a scale $\sim 30$ nm. Kohen
{\it et al}.\cite{koh03} measured the conductance of a contact between
gold and $\mathrm{Y_{1-x}Ca_xBa_2Cu_3O_{7-\delta}}$. The experimental
results were fitted with BTK-type\cite{BTK} (Blonder-Tinkham-Klapwijk)
model generalized to include the $d_{x^2-y^2}$ symmetry of the order
parameter,\cite{tan95} as well as a subdominant component of the order
parameter of $s$ or $d_{xy}$ symmetry. The authors of
Ref.~\onlinecite{koh03} concluded from their fits that the order
parameter on the superconductor side might have the time-reversal
symmetry breaking (TRSB) form $d+is$. This result was rather
surprising, since the contacts had predominantly [100] orientation and
very high transparency, in which case TRSB is usually not expected
(see e.g. Refs.~\onlinecite{mat95,fog97} and the reviews in
Refs.~\onlinecite{Kash_prog,LSW_rev}). Therefore, the authors
speculated that maybe there is an {\it unusual proximity effect},
where the TRSB state is induced by the proximity effect at a good
contact between a normal metal and $d_{x^2-y^2}$ superconductor with
[100] orientation. The authors of Ref.~\onlinecite{sha04} later
speculated, based on the size of the pseudogap seen in their STM
spectroscopy, that they might also have observed this type of new
symmetry breaking.

Motivated by these experiments, we set out to explore the proximity
effect in normal metal {\Large -} high-$T_c$ superconductor
structures. We do {\it not} find any evidence of a new symmetry
breaking proximity effect, within our model and its parameter
space. But, on the other hand, we are able to qualitatively explain
the pseudogap phenomenon seen in the latter experiment\cite{sha04} in
terms of a more conventional type of proximity effect where the
modulation in the local density of states is due to the interplay of
leaking superconducting correlations into the normal metal and
impurity scattering.

\section{Model and Methods}

\begin{figure}[t]
\includegraphics[width=8cm]{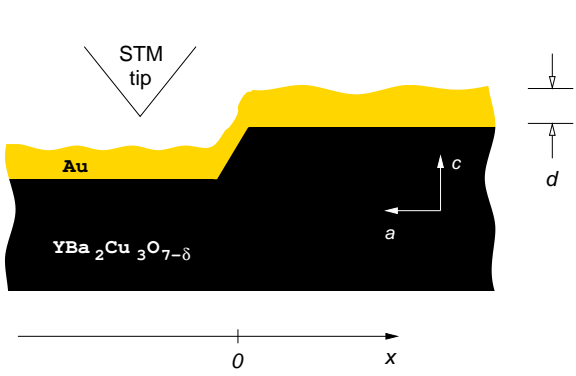}
\caption{We model the experimental setup by a half-infinite ($x\in
[-\infty,0]$) dirty normal metal in good contact with a [100] facet of
rather clean YBCO ($x\in [0,\infty]$). The LDOS in the normal metal
may be mapped out by STM.}\label{fig:NLayer_exp}
\end{figure}

\begin{figure}[t]
\includegraphics[width=8cm]{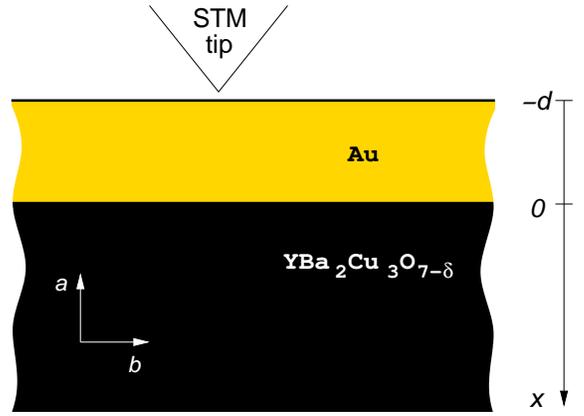}
\caption{The second model system is a more conventional proximity
overlayer, where the LDOS at the normal metal surface may be probed by
STM.}\label{fig:NLayer}
\end{figure}

We shall describe two types of systems: first, an NIS system where a
normal metal (N) is coupled to an [100] ($a$-axis) high-$T_c$
superconductor (S) facet through a high-transparency tunnel barrier
(I), as shown in Fig.~\ref{fig:NLayer_exp}. This setup roughly
corresponds to the experimental situation.\cite{sha04} We will also
present results for a [110]-contact. Second, an INIS system consisting
of an $a$-facet with a metal overlayer, shown in
Fig.~\ref{fig:NLayer}.

Our considerations will be applicable to rather general situations,
but in order to match the experimental conditions we restrict the
parameter space of our model. We focus on the low temperature region
$T=0.05T_c$, where $T_c$ is the superconducting transition temperature
($T=4.2K$ and $T_c=90K$ in the experiment). In the S-side we include,
besides the $d$-wave interaction, an attractive pairing interaction in
the $s$-wave channel with a critical temperature $T_{cs}=0.1T_c$,
which is similar in magnitude to previous estimates from
measurements\cite{cov97,gre00,Krupke,sha02} of TRSB in [110] oriented
tunnel junctions of YBCO. We note, however, that signatures of
subdominant pairing is not always seen
experimentally.\cite{alff_rapid,yeh01} In the N-side we include a
pairing interaction that can be either attractive, repulsive, or
zero. We assume that the gold metal is dirty, and include impurity
scattering in the self-consistent $t$-matrix
approximation,\cite{AGD,xu95} under the assumption that the impurities
scatter isotropically ($s$-wave scattering only) in the Born
limit. This corresponds to the Usadel approximation,\cite{usadel}
which is widely used to describe the proximity effect in contacts
involving low-$T_c$ superconductors. Since we are considering the
high-$T_c$ superconductors, which are clean materials with anisotropic
pairing, we can not use Usadel's scheme; instead we solve the full
Eilenberger equations\cite{eil68,SereneRainer} for the quasiclassical
Matsubara matrix Green's function $\hat g^M(\vp_f,\vR;\epsilon_n)$,
where $\vp_f$ is the Fermi momentum, $\vR$ is the spatial coordinate,
and $\epsilon_n=(2n+1)\pi T$ ($n$ integer) is the Fermion Matsubara
frequency:
\begin{equation}\label{eq:Q}\begin{split}
\left[ i\epsilon_n\hat\tau_3-\hat\sigma^M-\hat\Delta,\hat g^M \right]
+i\vv_f\cdot\nabla \hat g^M &= 0,\\
(\hat g^M)^2 &= -\pi^2.
\end{split}\end{equation}
Here is $\hat\tau_3$ the third Pauli matrix in Nambu space,
$\hat\Delta$ is the superconductor gap (pair potential), and
$\hat\sigma^M$ is the sum of the impurity elastic and spin flip self
energies $\hat\sigma^M=\hat\sigma^M_{imp}+\hat\sigma^M_{sf}$. We use
units $\hbar=k_B=1$, except in Table.~\ref{table:scales}.


The gap equation reads
\begin{equation}\label{eq:delta}
\hat\Delta(\vp_f,\vR) = T\sum_{\epsilon_n}^{|\epsilon_n|<\omega_c} 
\int d\vp_f'
\lambda(\vp_f,\vp_f',\vR) \hat f^M(\vp_f',\vR;\epsilon_n),
\end{equation}
where $\hat f^M$ is the off-diagonal part of the matrix Green's
function $\hat g^M$, $\omega_c$ is an energy cut-off, and $\int
d\vp_f...$ denotes a Fermi surface average. In our model system, we
assume that the interaction has a simple form
\begin{equation}\label{eq:lambda}
\lambda(\vp_f,\vp_f',\vR) = \left\{
\begin{array}{l}
\lambda_n,\; \mbox{N side},\\
\lambda_d\,\eta_d(\vp_f)\,\eta_d(\vp_f') + \lambda_s,\; \mbox{S side},
\end{array}\right.
\end{equation}
where $\lambda_d$ and $\lambda_s$ are the interaction strenghs in the
$d$-wave and $s$-wave channels, respectively, in the
superconductor. The basis function in the $d$-wave channel is taken to
have the form $\eta_d(\vp_f)=\sqrt{2}(\hat p_x^2-\hat p_y^2)$, $\int
d\vp_f \eta_d(\vp_f)=0$, $\int d\vp_f |\eta_d(\vp_f)|^2=1$. In the
normal metal for attractive interaction $\lambda_n>0$, while for
repulsive interaction $\lambda_n<0$. For attractive interaction, we
can eliminate the cut-off $\omega_c$ and the interaction $\lambda$ in
favor of the bare bulk superconducting transition temperature $T_c$ in
the usual way,\cite{Kieselmann} $\lambda^{-1}=\ln(T/T_c)+\sum_{n\geq
1} (n-0.5)^{-1}$. For repulsive interaction, this procedure obviously
fails, and we have a weak cut-off dependence. We take
$\omega_c=30T_c$, but we have checked that our results are not
qualitatively changed for other values of $\omega_c$.

With the pairing interaction in Eq.~(\ref{eq:lambda}), the order
parameter is
\begin{equation}\label{eq:deltaX}
\Delta(\vp_f,\vR) = \left\{
\begin{array}{l}
\Delta_n,\; \mbox{N side},\\
\Delta_d\,\eta_d(\vp_f) + \Delta_s,\; \mbox{S side},
\end{array}\right.
\end{equation}
where all amplitudes are in general complex quantities. We choose
$\Delta_d$ real, while the phases of $\Delta_n$ and $\Delta_s$
relative to $\Delta_d$ are found self-consistently. We denote the
maximum value of the $d$-wave component,
$\max\{\Delta_d\eta_d(\vp_f)\}=\sqrt{2}\Delta_d$, by $\Delta_0$.

The self energy from elastic impurity scattering has the
form\cite{AGD}
\begin{equation}\label{eq:sigma_imp}
\hat\sigma^M_{imp}(\vR,\epsilon_n) = \frac{1}{2\pi\tau_{imp}}
\int d\vp_f
\hat g^M(\vp_f,\vR;\epsilon_n),
\end{equation}
and we shall use the mean free path $\ell=v_f\tau_{imp}$ as the input
parameter, which takes the value $\ell_{n/s}$ in the N and S sides,
respectively. We also include spin-flip scattering through the self
energy\cite{RS_RMP86}
\begin{equation}\label{eq:sigma_sf}
\hat\sigma^M_{sf}(\vR,\epsilon_n) = \frac{1}{2\pi\tau_{sf}}
\int d\vp_f
\hat\tau_3
\hat g^M(\vp_f,\vR;\epsilon_n)
\hat\tau_3,
\end{equation}
where the mean free path $\ell_{sf}=v_f\tau_{sf}$ is always assumed to
be large compared to the elastic mean free path. The spin-flip process
will ultimately set an upper limit on how long range the proximity
effect is.\cite{bel96}

The interface connecting the normal metal side and the superconductor
side is described by Zaitsev's boundary condition,\cite{zai84} valid
for arbitrary transparency $\D(\vp_f)$. This model describes a wide
specularly scattering contact with translational invariance along the
interface. For simplicity we consider a two-dimensional system, with
quasiparticles moving within the $ab$-plane in the superconductor
which we also denote the $xy$-plane, with the $x$-axis extending
perpendicular to the interface (as in
Figs.~\ref{fig:NLayer_exp}-\ref{fig:NLayer}) and the $y$-axis lies
parallel to the interface. We assume circular Fermi surfaces in the
two sides and take the angular dependence of the interface barrier as
predicted by an interface $\delta$-function potential of strength
$H_b$, including the possibility of different Fermi velocities in the
two sides (see e.g. Ref.~\onlinecite{Kieselmann}). We assume that the
effective masses in the two sides are the same (for a discussion of
effective mass mismatch, see e.g. Ref.~\onlinecite{ash86}). For
Zaitsev's boundary condition the momentum parallel to the interface is
concerved, which means that the two different trajectory angles
$\theta_{n/s}$ of the two sides, measured relative to the $x$-axis,
are related as
\begin{equation}
v_{fs}\sin\theta_s = v_{fn}\sin\theta_n.
\end{equation}
We assume that the Fermi velocity is larger in the normal metal than
in the high-$T_c$ superconductor, $0<\alpha\equiv v_{fs}/v_{fn}\leq
1$. Quasiparticles travelling on trajectories with angles $\theta_n$
larger than a critical angle
\begin{equation}
\theta_c = \arcsin \alpha,
\label{thetaC}
\end{equation}
suffer total reflection. The interface transparency for angles
$\theta_n<\theta_c$ is
\begin{equation}
\D(\vp_f) = \frac{4v_nv_s}{(v_n+v_s)^2+4H_b^2},
\label{D_of_pf}
\end{equation}
where $v_{n/s}=v_{fn/s}\cos\theta_{n/s}$ are the projections on the
interface normal of the Fermi velocities in the two sides. We use the
model parameters $\alpha$ and $Z\equiv H_b/v_{fn}$ to parameterize the
barrier transparency. The transparency for perpendicular incidence is
denoted $\D_0=\alpha/[0.25(1+\alpha)^2+Z^2]$. We concentrate on the
high-transparency limit, with values given in the experimental
paper:\cite{koh03} $\D_0\sim 0.8-1.0$.

To compute the local density of states, we solve the retarded versions
of the above equations, which are obtained by changing the superscript
$^M$ to $^R$ and letting $i\epsilon_n\rightarrow \epsilon+i0^+$ in
Eqs.~(\ref{eq:Q}) and (\ref{eq:sigma_imp})-(\ref{eq:sigma_sf}). Here
is $i0^+$ an infinitesimally small positive imaginary number. To speed
up the convergence of the numerics we have kept this imaginary number
finite equal to $i10^{-3}$. The gap profiles obtained within the
Matsubara technique serve as input to the calculation of the retarded
Green's functions. The local density of states (LDOS) is defined as
\begin{equation}\label{eq:ldos}\begin{split}
\cN(\vR,\epsilon) &= \int d\vp_f \cN(\vp_f,\vR;\epsilon),\\
\cN(\vp_f,\vR;\epsilon) &= -\frac{\cN_f}{2\pi}
\mbox{Im}\left\{\mbox{Tr}\left[\hat\tau_3\,
\hat g^R(\vp_f,\vR;\epsilon)
\right]\right\},
\end{split}\end{equation}
where $\cN_f$ is the density of states at the Fermi level in the
normal state.

The above set of equations for the Green's functions and self-energies
are solved self-consistently with numerical methods. This can be done
quite efficiently with the Riccati parametrization
technique.\cite{nag93,sho95,ShelankovOzana,esc00}

\begin{table*}[t]
\begin{tabular}{lll}
\hline
\hline
 & NM  &\\
\hline
clean normal metal coherence length $\xi_{n0}=\frac{\hbar v_f}{2\pi k_B T}$ & $1$ $\mu$m @ 1K &\\
\hline
normal metal elastic mean free path $\ell_n=v_f\tau$ & $10-100$ nm &\\
\hline
dirty normal metal coherence length $\xi_n=\sqrt{\frac{\hbar D}{2\pi k_B T}}$ & $50-200$ nm @ $1$ K &\\
\hline
\hline
 & LTS & YBCO\\
\hline
superconductor gap size $\Delta_0$ & $\sim 1.5$ meV [\onlinecite{gup04}]& $20-25$ meV [\onlinecite{sha04,dam03}]\\
\hline
superconductor coherence length $\xi_{\Delta}=\frac{\hbar v_f}{\pi \Delta_0}$ & $140$ nm & $16$ \r{A} [\onlinecite{zhe94}]\\
\hline
elastic mean free path $\ell_s=v_f\tau$ & $10-100$ nm & $250-2500$ nm [\onlinecite{bon94,duf01}]\\
\hline
dirty superconductor coherence length $\xi_{s}=\sqrt{\frac{\hbar D}{2\Delta}}$ & 27-86 nm & not relevant ($\xi_{\Delta}\ll\ell_s$)\\
\hline
\hline

\end{tabular}
\caption{Order of magnitudes of some length scales relevant to the
proximity effect in normal metals (NM), low-$T_c$ superconductors
(LTS), and YBCO. The diffusion constant is defined as
$D=\frac{1}{3}v_f\ell$. For all estimates in NM and LTS we used
$v_f\sim10^6$ m/s [\onlinecite{AM}]. The cited LTS gap value is for Nb
[\onlinecite{gup04}]. The maximum $d$-wave gap amplitude is
$\Delta_0=\sqrt{2}\Delta_d$, and the magnitude given here is for the
$a$-axis direction. The short coherence length in YBCO is usually
estimated from the upper critical field in terms of Ginzburg-Landau
theory\cite{zhe94} $H_{c2}=\Phi_0/2\pi\xi^2$, where $\Phi_0=hc/2e$.}
\label{table:scales}
\end{table*}

\section{Results for the NIS system}

Previously, Ohashi\cite{oha96} made a thorough investigation of the
proximity effect in a clean system (mean free path
$\ell\rightarrow\infty$) between a normal metal with attractive or
repulsive $s$-wave interaction, in contact with a purely $d_{x^2-y^2}$
superconductor ($\lambda_s=0$). See also the paper by
Bruder.\cite{Bruder} The central result of Ref.~\onlinecite{oha96} is
that for the [100] orientation an $s$-wave pair potential is induced
in the N-side (for $\lambda_n\neq 0$), which also leads to spectral
changes. For the [110] orientation the gap amplitude vanishes. The
difference between the two orientations is easily understood by
symmetry arguments: the positive and negative lobes of the
$d_{x^2-y^2}$ order parameter contributes to the gap equation in the
N-side with equal weights (but with opposite signs) in the [110]
orientation and cancel exactly. This cancellation does not occur for
the [100] orientation, and an $s$ order parameter is induced. However,
these results does not answer all question raised in the
experiments.\cite{sha04,koh03} Also, the effects of impurities remain
unclear. In the following we therefore present an extensive
investigation of the proximity effect, including effects of impurity
scattering and subdominant pairing in the S-side.

\subsection{[100] orientation, $\lambda_n=0$}
\begin{figure}
\includegraphics[width=6.5cm,angle=-90]{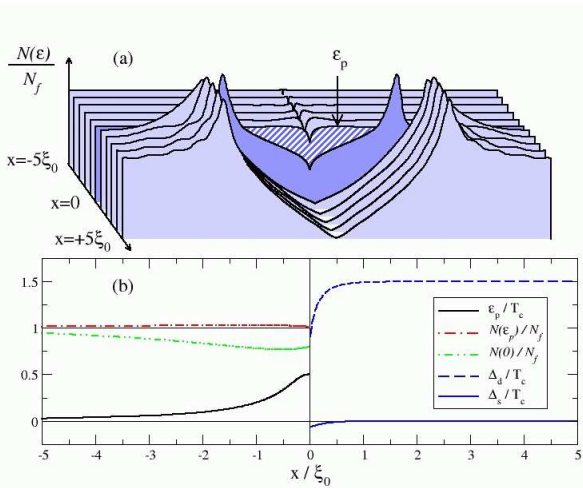}
\caption{The proximity effect at an [100] contact between a dirty
normal metal ($\ell_n=\xi_0$) and a rather clean $d_{x^2-y^2}$
superconductor ($\ell_s=25\xi_0$) with a subdominant $s$-wave
component ($T_{cs}=0.1T_c$). The pairing interaction in the normal
metal side is $\lambda_n=0$ (thus $\Delta_n=0$). The temperature is
$T=0.05T_c$, and the spin flip mean free path is $\ell_{sf}=100\xi_0$
in both sides. In (a) we show the density of states at $x=n\xi_0$
(integer $n=-5...5$) for $\epsilon\in[-4T_c,4T_c]$. There is a sudden
change at $x=0$ because of the interface backscattering for $\D_0=0.9$
($\alpha=1$, $Z=1/3$): the striped filled curve is at $x=0^-$ and the
dark filled is at $x=0^+$. The density of states is normalized to
$\cN_f$ so that it approaches $\rightarrow 1$ for $\epsilon\gg
T_c$. In (b) we show the spatial dependence of the peak
$\epsilon_p(x)$ in the LDOS in the normal metal, the corresponding
density of states $\cN(\epsilon_p)$, the density of states at the
Fermi level $N(0)$, and the pair potentials $\Delta_d$ and $\Delta_s$
in the superconductor.}\label{fig:NISnoint}
\end{figure}
In Fig.~\ref{fig:NISnoint} we present a representative picture of the
proximity effect at a [100] contact, in the absence of pairing
interaction in N-side ($\lambda_n=0$) but with a subdominant
interaction channel in the S-side ($T_{cs}=0.1T_c$). In part (a) we
show the LDOS, and in (b) we show the order parameter profiles and
some of the characteristic features in the LDOS. The parameters where
chosen such that the elastic mean free path in the superconductor is
$\ell_s=25\xi_0$, corresponding to the typical low-temperature
residual scattering rate $\tau\sim 1-10$ ps found in
experiments.\cite{bon94,duf01} The normal metal side is assumed dirty,
with $\ell_n=\xi_0=v_f/T_c\sim 10-100$ nm typical for metals in
general\cite{AM} and the Au overlayer in the experiment in
particular.\cite{sha04} Here we put $v_f\sim 10^6$ m/s in both sides,
for simplicity. Note that the superconductor coherence length
$\xi_{\Delta}=v_f/{\pi\Delta_0} \alt \xi_0/{2\pi}$ is much smaller
than the length scale $\xi_0=v_f/T_c$ we use. We summarize the
relevant length scales found in real materials in
Table~\ref{table:scales}.

If the normal metal was clean, there would not be any signatures of
the proximity effect in the LDOS\cite{oha96} (since the pair potential
is zero for $\lambda_n=0$). In contrast, in the dirty case presented
here we have clear signatures in the form of a pseudogap at low
energies, that decays into the normal metal on the thermal coherence
lenght scale $\xi_n=\sqrt{D/(2\pi T)}$, where $D=v_f\ell_n/3$ is the
diffusion constant in the normal metal. To quantify the effect we show
in Fig.~\ref{fig:NISnoint}(b) the spatial dependences of the pseudogap
peak position $\epsilon_p(x)$ (solid black line), the LDOS at the peak
position $\cN(\epsilon_p)$ (red dashed-dotted line), and the LDOS at
zero energy $\cN(0)$ (green dashed-double-dotted line), where it is
suppressed the most. The maximum deviation from the normal metal
density of states is about $25\%$, which is much smaller than the
$100\%$ effect predicted in low-$T_c$ $s$-wave
superconductors\cite{deu69,gue96,mou01,gol88,bel96,pil00,bel99} at
zero energy. This difference is due to the nodes in the
$d_{x^2-y^2}$-wave order parameter and the cancellation effects
between the positive and negative lobes, which substantially reduce
the influence of the unconventional superconductor on the LDOS in the
normal metal.

The presence of a proximity effect at the junction signals a non-zero
Fermi-surface average
\begin{equation}
\int d\vp_f f^R(\vp_f,\vR;\epsilon) \neq 0
\label{nonZeroFSA}
\end{equation}
in this region. This average is non-zero near the interface in both
sides. Its non-zero value is intimately related to the suppression of
the $d_{x^2-y^2}$ component.\cite{oha96,gol98,gol99} Since
$f^R(\vp_f,\vR;\epsilon)\propto \eta_d(\vp_f)$, a non-zero Fermi
surface average can only be obtained for a spatially dependent
$d_{x^2-y^2}$-wave gap in a self-consistent solution where
non-locality is taken into account. The average is also enhanced when
the back-scattering probability at the interface $\R_0=1-\D_0$ is
non-zero, since then the importance of high angle trajectories
$>\pi/4$ (for which the $d$-wave order parameter is negative) is
reduced compared to the low angle trajectories $<\pi/4$ (for which the
order parameter is positive).

In Fig.~\ref{fig:NISnoint} we couple back the non-zero Fermi surface
average of the off-diagonal Green's function in the superconductor
side by having a small subdominant pairing interaction
($\lambda_s>0$). We then obtain a non-zero subdominant order parameter
component $\Delta_s$ near the interface in the
superconductor. However, this component is always in phase with the
dominant $d$-wave, and time-reversal symmetry is always conserved at
the [100]-contact in the parameter space we have explored (mean free
paths ranging from the clean limit $\ell_n\rightarrow\infty$ to the
dirty limit $\ell_n\ll\xi_{n0}$ with a wide variety of barrier
transparencies and subdominant pairing interactions). The effects of
the $s$-wave gap on the local density of states is marginal.

\subsection{[100] orientation, $\lambda_n<0$}
\begin{figure}[t]
\includegraphics[width=6.5cm,angle=-90]{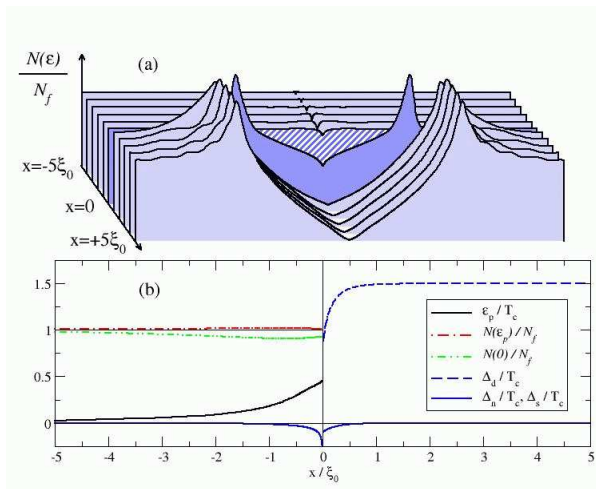}
\caption{The same as in Fig.~\ref{fig:NISnoint}, but with repulsive
pairing interaction in the normal metal side $\lambda_n=-5$, which
result in an induced pair-potential $\Delta_n$ in the normal
metal.}\label{fig:NISwithint}
\end{figure}
In Fig.~\ref{fig:NISwithint} we present the proximity effect at a
[100] contact, for the case of a repulsive pairing interaction in the
N-side. We have in mind that the counter electrode in the
experiment\cite{sha04} was Au which is not a superconductor and might
have a repulsive effective pairing interaction. The $d_{x^2-y^2}$-wave
pairing correlations leaks over to the N-side, which leads to the
non-zero Fermi-surface average in Eq.~(\ref{nonZeroFSA}). That average
is coupled back though the repulsive pairing interaction $\lambda_n$,
which leads to a non-zero pair potential $\Delta_n$, that decays into
the bulk normal metal. The subdominant component $\Delta_s$ on the
S-side is not qualitatively affected by the induced pair potential on
the N-side. The opposite is also true, the induced $\Delta_n$ is not
qualitatively affected by the presence ($\lambda_s\neq 0$) or absence
($\lambda_s=0$) of the component $\Delta_s$. Thus, the proximity
induced pair potential on the normal metal side is mainly due to the
dominant $d_{x^2-y^2}$ component. We also note that the real
combination $d+s$ is always favored in the system, as in the case
$\lambda_n=0$ above.

As for the LDOS, the induction of the gap amplitude $\Delta_n$ changes
the density of states near the contact. There are new low-energy
states, which are formed because of the sign change of the order
parameter field for quasiparticles travelling on low-angle
trajectories $<\pi/4$ which connect the negative gap $\Delta_n$ on the
N-side with the positive lobe of the $d_{x^2-y^2}$ order parameter on
the S-side. This is analogous to the zero-energy bound states formed
at a [110] surface of a $d_{x^2-y^2}$ superconducor,\cite{hu94} and
was also discussed in the study of the superclean case.\cite{oha96} In
our case, the impurity scattering broadens the states considerably.
Thus, compared with the case $\lambda_n=0$ above, the LDOS peak
position $\epsilon_p(x)$ is drawn closer to the Fermi surface. At the
same time, the overall modulation of the LDOS in the normal metal is
reduced compared to the case without the pairing interaction. Thus,
contrary to our first expectations, an induced non-zero pair potential
$\Delta_n$ in the normal metal reduces the proximity effect in the
LDOS, as compared to the case $\Delta_n\equiv 0$. This is a unique
feature of the dirty system.

\subsection{[100] orientation, $\lambda_n>0$}
\begin{figure}[t]
\includegraphics[width=6.5cm,angle=-90]{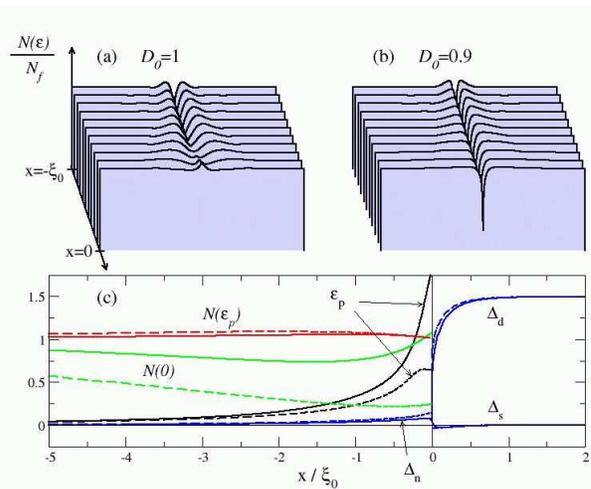}
\caption{Proximity effect at a [100] contact with a normal metal with
attractive pairing interaction ($T_n=0.01T_c\,<\,T=0.05T_c$), for high
barrier transparencies $\D_0=1$ ($\alpha=1$, $Z=0$) in (a) and
$\D_0=0.9$ ($\alpha=1$, $Z=1/3$) in (b). All other model parameters
are the same as in Figs.~\ref{fig:NISnoint}. Only the density of
states in the normal metal at $x\in[-\xi_0,0]$ (in steps of
$0.1\xi_0$) for $\epsilon\in[-4T_c,4T_c]$ is shown in (a)-(b). The
solid (dashed) lines in (c) refer to the contact with $\D_0=1$
($\D_0=0.9$).}
\label{fig:NIS_Tn0.01}
\end{figure}
For the case of attractive interaction in the N-side, $\lambda_n$ is
positive which results in a sign change of the order parameter
$\Delta_n$ compared to the repulsive interaction case considered above
(see Fig.~\ref{fig:NIS_Tn0.01}). At the same time the order parameter
amplitude becomes larger and decays slower compared to the repulsive
interaction case (analogous results were also found for the clean case
in Ref.~\onlinecite{oha96}). As for the LDOS, the weight of the
low-energy states is larger because of the larger pair potential,
although the trajectories contributing to these states are at high
angles $>\pi/4$. For a fully transparent interface, the states are
centered at zero energy, Fig.~\ref{fig:NIS_Tn0.01}(a). For non-zero
back-scattering at the junction, the states are split, see
Fig.~\ref{fig:NIS_Tn0.01}(b), where the split size depends on the
interface transparency ($\D_0=0.9$ in the figure). The total effect of
the proximity effect is enhanced for the case of attractive paring
interaction in the normal metal, compared to the case without
interaction in Fig.~\ref{fig:NISnoint}. The LDOS variations can be as
high as $~80\%$ for our choice of parameters, see
Fig.~\ref{fig:NIS_Tn0.01}(c).

Given the fact that there are low-energy states we are tempted to
think that a TRSB combination $d_{x^2-y^2}+is$ could be formed (at
least for $\D_0\equiv 1$ for which the split-effect is absent),
instead of the real $d_{x^2-y^2}+s$ combination found above. If this
was the case, the energy of the system might be lowered, similar to
what happens at [110] surfaces or Josephson
junctions.\cite{sig98,FY,LSW_prb00,ami01} However, for the proximity
contact we were never able to stabilize a TRSB state within the
parameter space we have considered. We understand this result as due
to a phase locking between the $s$-wave gaps in the N and S sides, to
the dominant $d_{x^2-y^2}$, that prevents a relative phase $\pi/2$
between any two components to be formed. Compared to the Josephson
junctions case, the two sides of the proximity contact are not
independent: there is no bulk order parameter on the N-side that makes
it possible to establish a phase difference $\pi/2$ over the
junction. As for the subdominant $s$-wave order parameter on the
S-side, it is due to the spatially varying $d_{x^2-y^2}$ order
parameter, and is not induced by the bound state as in the [110]
surface case. We also note that spontaneous symmetry breaking in the
form of a paramagnetic
instability\cite{Higashitani,fau99,bkk_prb00,LSW_prb00,ShelankovOzana}
is suppressed by impurity scattering and normal back-scattering at the
junction.

\subsection{[110] orientation}
\begin{figure}
\includegraphics[width=6.5cm,angle=-90]{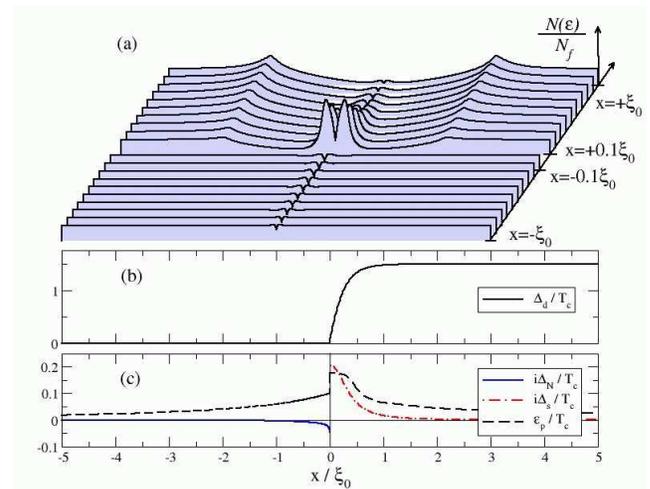}
\caption{The proximity effect at a [110] contact with $T_{cs}=0.1T_c$.
A subdominant component developes near the surface in the TRSB
combination $d+is$ at the S-side. For intermediate transparencies,
here $\D_0=0.3$ ($\alpha=1$, $Z^2=7/3$), the subdominant component
induces an $is$-wave pair potential at the N-side ($\lambda_n=-5$). We
have $T=0.05T_c$, $\ell_n=\xi_0$, $\ell_s=25\xi_0$, and
$\ell_{sf}=100\xi_0$.}\label{fig:NIS_110}
\end{figure}
We now turn to the [110] orientation. In previous studies of the clean
system,\cite{oha96} no gap amplitude was found in the normal metal
side for reasons of symmetry. As for the LDOS in the N side, it is
always unchanged in a clean system when the gap amplitude is zero. For
$\lambda_n\equiv 0$, we find no signatures of the proximity effect in
the LDOS in the dirty case either, which also follows from symmetry
arguments. However, when we include a subdominant component in the
superconductor side at a contact with [110] orientation, it will be
formed $\pi/2$ out of phase with the dominant, and we have a TRSB
state \cite{mat95,fog97} (this is opposite to the [100] case
above). We note that some backscattering is necessary at the junction,
i.e. if the transparency $\D_0\equiv 1$ there are no zero-energy
states and no TRSB. As for the proximity effect in the N-side, high
transparency is favorable. Thus, we concentrate on a window of
intermediate transparencies, in order to establish a proximity effect
between a TRSB $d_{x^2-y^2}+is$ state at a [110] interface to a normal
metal, see Fig.~\ref{fig:NIS_110}. Clearly, the established $is$-part
in the S-side, leads to a proximity induced $is$ component also in the
N-side that decays into the bulk normal metal, although the effect is
rather small and short range. The details of the window of
transparencies depend on the mean free paths, as well as the
interaction strengths. Typical values where there are TRSB and at the
same time visible effects in the LDOS on the N-side are $\D_0\sim
0.1-0.5$, for our choice of parameters. The double peak structure on
the S-side, see the (a)-part of the figure, decays on the coherence
length scale, together with the subdominant component, red-dashed line
in (c). Both this peak structure, and the peak structure in the LDOS
at the N-side, [black dashed line in (c)], are related to, but not
direct images of, the $s$-wave pair potentials $i\Delta_s$ and
$i\Delta_n$ because of the importance of the impurity self-energy.

These results have the potential to serve as a probe of TRSB at [110]
interfaces: as the temperature is decreased, the spectrum in the
normal metal overlayer remains unchanged (equal to $\cN_f$ for all
energies) until the TRSB transition is reached, below which a small
pseudogap develops as shown in Fig.~\ref{fig:NIS_110}. This probe is
however not very quantitative and will not give direct information
about the size of $\Delta_s$, because of the masking effect of the
impurity scattering. Rather, it would serve as a {\it yes/no}
experiment, as to the existence or absence of TRSB at low
temperatures.

The drawback of the above considerations is that they rely heavily on
symmetry arguments. Thus, to a much higher degree than other results
in this paper, non-specular interface
scattering\cite{MS1,BSB,luc01,asa01,asa02,nag04} could be detrimental.

Another caveat is the possibility of $d$-wave pairing correlations
(attractive or repulsive) in the normal metal. An induced $d$-wave
pair potential leads to spectral changes in the normal metal for the
[110] orientation also in the absence of TRSB, thereby spoiling the
symmetry argument. The LDOS is enhanced at low energy by the formation
of low-energy Andreev states broadened by impurity scattering, if the
$d$-wave pairing correlations are oriented as [110] relative to the
interface. This is similar to what happens at Josephson junctions
between two $d$-wave superconductors.\cite{LSW_rev} However, it seems
unlikely that spatially homogeneous $d$-wave pairing correlations with
a well-defined orientation relative to the interface could exist in a
normal metal such as Au.

However, these complications are temperature-independent and should
not appear suddenly at the low-temperature TRSB transition. Thus, a
smoth but rather sudden change in the normal metal spectrum near the
interface as a function of temperature can be a signal of TRSB.

\section{Results for the INIS system}

\begin{figure}[t]
\includegraphics[width=8cm]{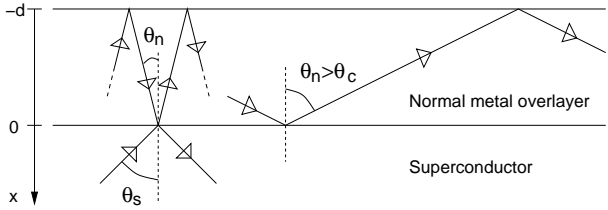}
\caption{Quasiparticles travelling on high-angle trajectories
$\theta_n>\theta_c$ suffer total reflection at the NIS interface
because of the Fermi velocity mismatch $v_{fn}>v_{fs}$ (the trajectory
to the right). On other trajectories $\theta_n<\theta_c$,
quasiparticles are partially reflected in accordance with
Eq.~(\ref{D_of_pf}).}
\label{fig:NLayer_trajectories}
\end{figure}
Let us now study the change in the density of states in a normal metal
overlayer in good contact with a [100] facet, see
Fig.~\ref{fig:NLayer}. Because of the absence of a bulk boundary
condition in the normal metal overlayer, we solve Eq.~(\ref{eq:Q})
iteratively until self-consistency is reached in the overlayer, under
the asumption of specular reflection at the vacuum-normal metal
surface (at $x=-d$). Including the possibility of a Fermi velocity
mismatch, we have two types of trajectories when we solve
Eq.~(\ref{eq:Q}) depending on if the trajectory angle $\theta_n$ is
larger or smaller than the critical angle for total inner reflection
in Eq.~(\ref{thetaC}). We show the trajectories in
Fig.~\ref{fig:NLayer_trajectories}. Note that when we draw the
trajecories in Fig.~\ref{fig:NLayer_trajectories} we do not imply that
quasiparticles move ballistically. Rather, the lines denote
trajectories along which we solve Eq.~(\ref{eq:Q}). Quasiparticles
diffuse, which is described by solving the impurity self-energies in
Eqs.(\ref{eq:sigma_imp})-(\ref{eq:sigma_sf}) self-consistently with
the Green's function in Eq.~(\ref{eq:Q}).

Because of the severe isotropization effect caused by impurity
scattering, we do not expect non-specular scattering to drastically
alter our results. In fact, similar results for the LDOS in the
superconductor as we find was already reported by Golubov and
Kupriyanov.\cite{gol98,gol99} They used the dirty normal metal
overlayer as a model for a diffusive boundary condition for the
superconductor. In their case, however, the dirty layer thickness was
assumed small, $d\ll\sqrt{\xi_0\ell_n}$. Here we relax this assumption
and study in more details the effect of the superconductor in the
normal metal.

\begin{figure}[t]
\includegraphics[width=6.5cm,angle=-90]{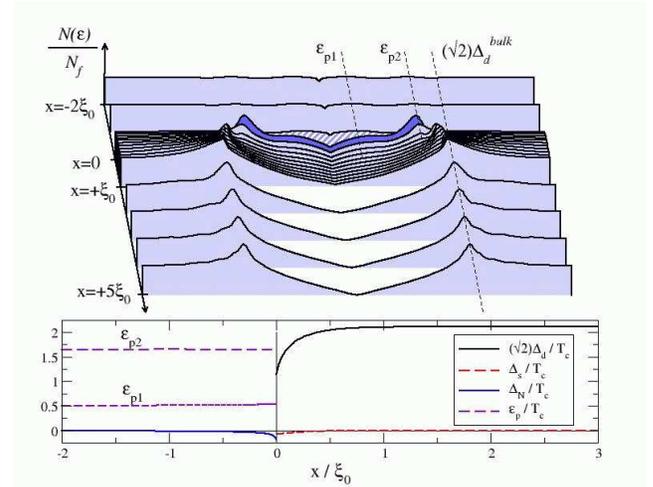}
\caption{The proximity effect at a [100] facet with a $d=2\xi_0$ thick
normal metal overlayer. We have $T=0.05T_c$, $T_{cs}=0.1T_c$,
$\lambda_n=-5$, $\ell_n=v_{fn}/T_c=2v_{fs}/T_c=2\xi_0$,
$\ell_s=25\xi_0$. The Fermi velocity mismatch is $v_{fn}=2v_{fs}$
($Z=0$), which leads to $\D(\theta_n=0)\approx 0.9$ and
$\theta_c\approx 30^o$.}
\label{fig:INIS}
\end{figure}
We present typical results for the order parameters and the LDOS for a
[100] contact in Fig.~\ref{fig:INIS}. We find that the LDOS is almost
independent of coordinates in the normal metal, but depends on the
thickness of the overlayer. There are quasibound states, seen as peaks
in the LDOS, reflecting interference between electrons and holes in
the dirty normal metal in contact with the
superconductor.\cite{LSW_rev} The energies of these states depends on
the thickness of the overlayer. In a normal metal in good contact with
an $s$-wave superconductor, there is a minigap $E_g$ of the order of
the Thouless energy $E_{th}=D/d^2$ ($D=\frac{1}{3}v_f\ell$), below
which the density of states vanish (for a discussion see
e.g. Ref.~\onlinecite{pil00} and references therein). Because of the
nodes of the $d_{x^2-y^2}$ order parameter and impurity scattering,
this effect is removed and only a suppression of the low-energy
density of states remains. As in the NIS case discussed in the
previous section, the LDOS in the normal metal mainly reflects the
influence of the impurity self-energy and does not contain direct
information about the induced pair potential $\Delta_n$ for
non-vanishing interaction $\lambda_n\neq 0$.

\begin{figure}[t]
\includegraphics[width=6.5cm,angle=-90]{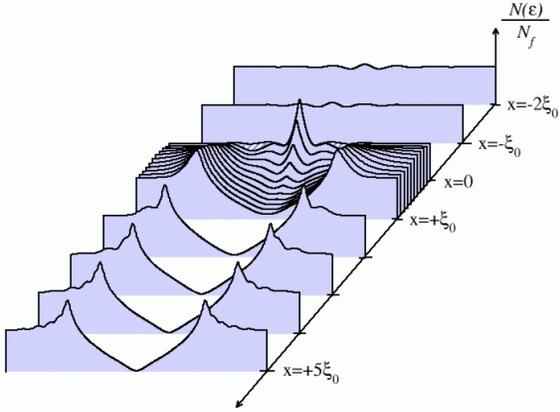}
\caption{The same as in Fig.~\ref{fig:INIS}, but for the [110]
orientation. For these model parameters only a $d$-wave order
parameter exist. The peak positions in the normal metal are at
$\epsilon_p=\{0,0.96T_c,1.98T_c\}$.}
\label{fig:INIS_110}
\end{figure}
In Fig.~\ref{fig:INIS_110} we show typical results for the [110]
contact. The peaks in the LDOS in the overlayer are
shifted,\cite{LSW_rev} as compared with the [100] contact. This
reflects the formation of the zero-energy states.\cite{hu94,LSW_rev}
For lower transparencies, the TRSB state can be favorable, which leads
to a split of the zero-energy peak, similar to the situation in
Fig.~\ref{fig:NIS_110}.

\begin{figure}[t]
\includegraphics[width=8cm]{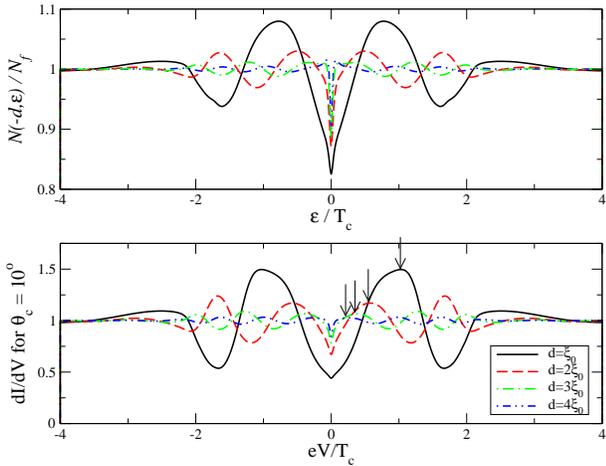}
\caption{(a) The surface local density of states and (b) the
conductance as it would be measured by STM according to
Eq.~(\ref{eq:dIdV}) for a $\theta_c^{tip}=10^o$ wide tunnel cone, for
several different thicknesses of the normal metal overlayer ranging
from $d=\xi_0$ to $d=4\xi_0$. The mean free path is $\ell_n=2\xi_0$
and total isotropization by the impurities has not yet occured which
leads to an enhancement of the modulation in the surface density of
states as seen in the STM conductance [note the different scales in
(a) and (b)]. The arrows indicate how the peak positions move toward
the Fermi level for increased overlayer thickness. The other model
parameters are the same as in Fig.~\ref{fig:INIS}. Note that the red
dashed curves corresponds to the data in Fig.~\ref{fig:INIS}.}
\label{fig:dIdV}
\end{figure}
In Fig.~\ref{fig:dIdV} we plot the surface LDOS and the conductance at
the [100] contact as it would be measured by an STM tip, as a function
of layer thickness. For simplicity we model the tunnel barrier between
the tip and the overlayer by a tunnel cone of width
$2\theta_{cone}^{tip}$, with transparency $\D_0^{tip}$ for angles
within the cone and transparency $0$ for angles outside. The
conductance is related to the angle resolved surface local density of
states $\cN(x=-d,\theta_n;\epsilon)$ as
\begin{equation}
\frac{dI}{dV}(V) = \frac{1}{R_n^{tip}}
\int_{-\theta_c^{tip}}^{\theta_c^{tip}} d\theta_n
\D_0^{tip}
\frac{\cN(-d,\theta_n;eV)}{\cN_f},
\label{eq:dIdV}
\end{equation}
where $R_n^{tip}$ is the resistance of the tip-sample contact in the
normal state (we normalize $\int d\theta_n\, \D^{tip}(\theta_n) =
1$). For intermediate overlayer thicknesses, total isotropisation of
the LDOS by the impurities might not have occured, and a small tunnel
cone can enhance the structures in the LDOS as seen by the STM, since
trajectories close to the interface normal is selected.

\section{Summary and Conclusions}

We have studied the proximity effect in good contacts between
$d_{x^2-y^2}$ superconductors and normal metals. We have investigated
the effects of impurity scattering, a subdominant component of
$s$-wave symmetry in the superconductor side, and a proximity induced
order parameter in the N side. For the [100] orientation, we always
find a real combination $d_{x^2-y^2}+s$ in the superconductor. At the
same time, the N side gap amplitude is phase locked to the $d$-wave
order parameter on the S-side. On the other hand, for a [110] contact
with intermediate values of the transparency, a TRSB $d_{x^2-y^2}+is$
order parameter on the S side, induces an $is$ gap amplitude also on
the N side. We note that the subdominant $is$ component in the S side
is not due to the proximity effect; rather, it is due to the formation
of the bound state at the [110] interface in the usual
way.\cite{mat95,fog97} The signatures of the proximity effect at [110]
contacts could serve as fingerprints of TRSB, with the caveat of
possible breaking of the ``selection rule'' by e.g. non-specular
scattering or $d$-wave pairing correlations in the normal metal.

We finish with a comparison with the experiments.\cite{koh03,sha04} We
can not find evidence of the unusual symmetry breaking proximity
effect at [100] contacts that was proposed in Ref.~\onlinecite{koh03}.
The proximity induced pair potential in the normal metal is due to the
suppression of the $d$-wave order parameter and no non-trivial
relative phase develops. However, a serious comparison with the
experiment of Kohen {\it et al.}\cite{koh03} can not be done since
that was a transport experiment where the conductance was measured. At
the same time, the analysis\cite{koh03} in terms of a BTK-type theory
with postulated pair potentials and no considerations of real
non-equilibrium effects, rather than self-consistently computed
quantities, might be too simplistic. Especially when the contact has
high transparency. For example, a dirty normal metal (instead of a
clean metal as in the BTK theory) in contact with a $d$-wave
superconductor can lead to quite different results for the
conductance, as shown theoretically recently.\cite{tan03,tan04} Those
considerations were limited to point contacts with the superconductor
a purely $d$-wave ($\lambda_s=0$), super-clean
($\ell_s\rightarrow\infty$), unperturbed reservoir.

On the other hand, our results is in {\it qualitative} agreement with
the experimental results of Sharoni {\it et al.}\cite{sha04} The
modulations of the density of states consists of a suppression of the
density of states at the Fermi level (but not to zero), with a quite
well defined peak position $\epsilon_p$. These structures decay into
the normal metal on the scale $\xi_n$. However, an {\it unambiguous
quantitative} fit between theory and data is hard to make for several
reasons. On the theory side, even within our simple model, the local
density of states is sensitive to many parameters, such as the mean
free paths, the interface resistance, the Fermi velocities, as well as
pairing interaction strengths. The choice of parameters is not
unique. On the experimental side, only a limited number of spectra
were shown, although many points $\epsilon_p(x)$ were reported. In
particular, the spectra for small distances were not shown, for which
$\epsilon_p$ were reported to be spectacularly large $\sim 15$ meV
$\sim 0.75\Delta_0$, which is hard to reproduce in theory
($\epsilon_p\sim 0.25\Delta_0$ in
Figs.~\ref{fig:NISnoint}-\ref{fig:NIS_110}). We note, however, that
similar (but not as large) discrepancies between theory and
experiments were recently reported\cite{gup04} also for low-$T_c$
contacts (Nb-Au). Although in that case, the discrepancy was the
largest for large distances. Another problem is that the modulation in
the LDOS appears to be larger than we find (unless the effective
pairing interaction in Au is attractive which seems unlikely). In
particular $\cN(\epsilon=0)$ is suppressed more than we find for the
NIS system. On the other hand, the experimental situation was not
clean cut: a mixture of the NIS and INIS systems appears to be
relevant. For large (small) distances from the $a$-facet, the setup
appears to be NIS-like (INIS-like). It is also unclear where along the
$x$-axis in our Fig.~\ref{fig:NLayer_exp} the spectra were actually
recorded. From their Figs.~1-2 it appears that the spectra were
actually taken at points corresponding to $x>0$ in our
Fig.~\ref{fig:NLayer_exp}, i.e. into the YBCO $c$-axis.

Despite these complications, it is quite plausible that the
experimental results could be explained by the more conventional type
of proximity effect we have discussed in this paper, rather than an
exotic proximity effect that involves symmetry breaking.

\begin{acknowledgments}
I would like to thank J.A.~Sauls for encouragements to finish this
work, and for pointing out the potential importance of repulsive
$d$-wave pairing correlations in the normal metal. Financial support
was provided by the Swedish Foundation for International Cooperation
in Research and Higher Education (STINT) and the Wenner-Gren
foundations.
\end{acknowledgments}

\end{document}